\documentclass[draft,preprint,aps,
showpacs
]{revtex4}

\usepackage{epsfig}

\usepackage{graphicx}
\usepackage{bm}

\input{epsf}

\newcommand{\be}{\begin{equation}}
\newcommand{\ee}{\end{equation}}
\newcommand{\bea}{\begin{eqnarray}}
\newcommand{\eea}{\end{eqnarray}}
\newcommand{\nn}{\nonumber \\}
\newcommand{\Bk}{{\bf k}}

\begin{document}

\preprint{Guchi-TP-012}
\date{\today%
}
\title{
Shape of Deconstruction
}

\author{Kiyoshi Shiraishi}
\email{shiraish@po.cc.yamaguchi-u.ac.jp}
\affiliation{Graduate School of Science and Engineering, Yamaguchi University, 
Yoshida, Yamaguchi-shi, Yamaguchi 753-8512, Japan}
\affiliation{Faculty of Science, Yamaguchi University,
Yoshida, Yamaguchi-shi, Yamaguchi 753-8512, Japan}

\author{Kenji Sakamoto}
\email{b1795@sty.cc.yamaguchi-u.ac.jp}
\affiliation{Graduate School of Science and Engineering, Yamaguchi University, 
Yoshida, Yamaguchi-shi, Yamaguchi 753-8512, Japan}

\author{Nahomi Kan}
\email{b1834@sty.cc.yamaguchi-u.ac.jp}
\affiliation{Graduate School of Science and Engineering, Yamaguchi University, 
Yoshida, Yamaguchi-shi, Yamaguchi 753-8512, Japan}

\begin{abstract}

We construct a six-dimensional Maxwell theory using a latticized
extra space, the continuum limit of which is a shifted torus recently
discussed by Dienes.
This toy model exhibits the correspondence between continuum theory
and discrete theory, 
and give a geometrical insight to theory-space model building.

\end{abstract}



\maketitle


\section{Introduction}

Though the idea that the dimension of the space-time may be more than four
is very old, the possible existence of the extra space becomes
a serious subject of theoretical and phenomenological study on
unified theories only recently.
In the view of the four-dimensional space-time,
every kind of the field in higher dimensions has the
corresponding Kaluza-Klein (KK) spectrum, if the extra space is compact.
The mass of the excited state is proportional to the
inverse of the size of the extra space.

Recently Dienes claimed that the KK spectrum 
depends on the shape as well as the volume of the 
extra two-torus\cite{Dienes,Dienes0}.%
\footnote{See also \cite{DM1} (for a three-torus), \cite{DM2}
(for the closed string theory) and \cite{CC} (for a noncommutative
field theory).}
The consideration of the shape moduli of the manifold may alter 
experimental bounds on the compactification scale.


On the other hand, 
there is another scheme to reexamine the concept of space,
which is known as deconstruction\cite{ACG,HPW}.
Suppose a number of copies of a four-dimensional theory.
Further add a new set of fields, linking pairs of these individual
``sites'' in the theory
space. The resulting whole theory may (or may not) be equivalent to a 
higher-dimensional theory with discretized, or, latticized extra dimensions.
Evidence or signal for this kind of deconstructed dimension
would be the discovery of a finite and specific ``KK spectrum'' in high energy
experiments.
 
A simple latticized model with a large number of sites
in ``two-dimensional'' theory space\cite{Lane}
resembles a continuum compactification.
It may be known that the discrete version of the shifted torus which Dienes 
examined 
can be (de)constructed naturally. Unfortunately, little attention has been paid to
the detail of such formulation. 
When we wish to distinguish between the discrete and continuum theories
by mean of future experiments, we should prepare the several
detailed templates of the mass spectra.


In this paper, we analyze the six-dimensional $U(1)$ gauge theory with a
latticized torus. We explicitly show the mass spectrum, which becomes the
KK spectrum of the continuum theory in the limit of the large number of sites.
Our toy model exhibits a clear correspondence between continuum theory
and discrete theory and may give a physical and geometrical insight to model
building.

In Sec.~\ref{sec:2}, we study the mass spectrum which comes from
 the $U(1)$ gauge field on a shifted lattice.
The one-loop effective potential for the Wilson line in
scalar QED is discussed in Sec.~\ref{sec:3}.
We close with Sec.~\ref{sec:4}, where summary and conclusion are given.

\section{shifted lattice}
\label{sec:2}

The discretization of a two-torus is realized by a doubly
periodic lattice with $N_u\times N_v$ sites.
The lagrangian we first consider is%
\footnote{
The quartic term such as
$U_{k\ell}V_{k,\ell+1}U_{k+1,\ell+h}^*V_{k\ell}^*$ may be 
included\cite{ACG,HPW,Lane} or
generated by radiative correction\cite{GW}; we omit such a term in this paper, for
mass spectra of vector bosons are unchanged by the term.}

\be
{\cal L}_A=\sum_{k=1}^{N_v}\sum_{\ell=1}^{N_u}\frac{1}{g^2}
\left[-\frac{1}{4}F_{k\ell}^{\mu\nu}F_{k\ell~\mu\nu}-
(D^{\mu}U_{k\ell})^{\dagger}D_{\mu}U_{k\ell}-
(D^{\mu}V_{k\ell})^{\dagger}D_{\mu}V_{k\ell}-V(|U_{k\ell}|,|V_{k\ell}|)
\right]\, ,
\label{lag}
\ee
where $g$ is a gauge coupling,
$F_{k\ell}^{\mu\nu}=\partial^{\mu}\tilde{A}_{k\ell}^{\nu}-
\partial^{\nu}\tilde{A}_{k\ell}^{\mu}$
and $\mu, \nu=0, 1, 2, 3$. The link fields $U_{k\ell}$ and $V_{k\ell}$ are
transformed as
\be
U_{k\ell}\rightarrow W_{k\ell}U_{k\ell}W_{k,\ell+1}^*\, ,\quad
V_{k\ell}\rightarrow W_{k\ell}V_{k\ell}W_{k+1,\ell+h}^*\, ,
\ee
under $U(1)^{N_uN_v}$ ($|W_{k\ell}|=1$ for any $k, \ell$).
Here $h$ is an integer. For $h=0$, a similar model has been analyzed
by Lane\cite{Lane} (see FIG.~1).

\begin{figure}[htb]
\centering
\unitlength=1mm
\begin{picture}(60,80)
\multiput(21,20)(10,0){3}{\vector(1,0){8}}
\multiput(21,30)(10,0){3}{\vector(1,0){8}}
\multiput(21,40)(10,0){3}{\vector(1,0){8}}
\multiput(21,50)(10,0){3}{\vector(1,0){8}}
\multiput(21,60)(10,0){3}{\vector(1,0){8}}
\multiput(20,21)(0,10){5}{\vector(0,1){8}}
\multiput(30,21)(0,10){5}{\vector(0,1){8}}
\multiput(40,21)(0,10){5}{\vector(0,1){8}}
\multiput(20,20)(10,0){3}{\circle{2}}
\multiput(20,30)(10,0){3}{\circle{2}}
\multiput(20,40)(10,0){3}{\circle{2}}
\multiput(20,50)(10,0){3}{\circle{2}}
\multiput(20,60)(10,0){3}{\circle{2}}
\end{picture}
\begin{picture}(60,80)
\multiput(20.7,20.7)(10,0){3}{\vector(1,1){8.6}}
\multiput(20.7,30.7)(10,0){3}{\vector(1,1){8.6}}
\multiput(20.7,40.7)(10,0){3}{\vector(1,1){8.6}}
\multiput(20.7,50.7)(10,0){3}{\vector(1,1){8.6}}
\multiput(20.7,60.7)(10,0){3}{\vector(1,1){8.6}}
\multiput(20,21)(0,10){5}{\vector(0,1){8}}
\multiput(30,21)(0,10){5}{\vector(0,1){8}}
\multiput(40,21)(0,10){5}{\vector(0,1){8}}
\multiput(20,20)(10,0){3}{\circle{2}}
\multiput(20,30)(10,0){3}{\circle{2}}
\multiput(20,40)(10,0){3}{\circle{2}}
\multiput(20,50)(10,0){3}{\circle{2}}
\multiput(20,60)(10,0){3}{\circle{2}}
\end{picture}
\caption{%
A schematic view of the two-types of the latticized two-tori. 
Since the diagrams are similar to FIG.~4 in [7], 
labels for each site and link are omitted. The left graph
corresponds to the case with $h=0$, while the right one corresponds to
the case with $h=1$.}
\label{fig1}
\end{figure}
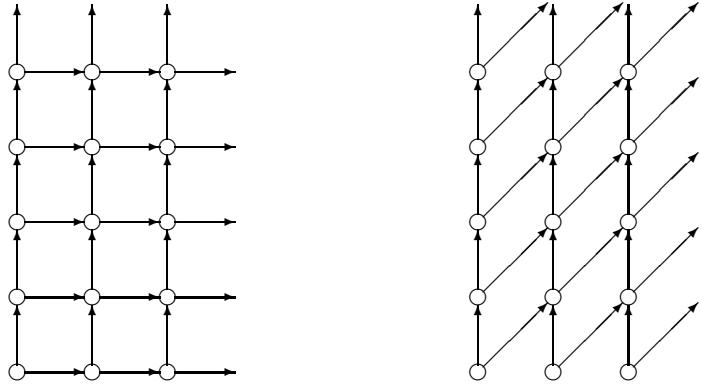

Then the covariant derivatives are
\bea
D^{\mu}U_{k\ell}&=&\partial^{\mu}U_{k\ell}-i\tilde{A}_{k\ell}^{\mu}U_{k\ell}+
iU_{k\ell}\tilde{A}_{k,\ell+1}\, ,\\
D^{\mu}V_{k\ell}&=&\partial^{\mu}V_{k\ell}-i\tilde{A}_{k\ell}^{\mu}V_{k\ell}+
iV_{k\ell}\tilde{A}_{k+1,\ell+h}\, .
\eea

We assume that the potential $V$ forces each $|U_{k\ell}|$ into 
$f_u/\sqrt{2}$ and each $|V_{k\ell}|$ into $f_v/\sqrt{2}$.
Then $U_{k\ell}$ and $V_{k\ell}$ are expressed as
\be
U_{k\ell}=\frac{f_u}{\sqrt{2}}
\exp\left(i\tilde{\chi}_{k\ell}/f_u\right)\, ,\qquad
V_{k\ell}=\frac{f_v}{\sqrt{2}}
\exp\left(i\tilde{\sigma}_{k\ell}/f_v\right)\, .
\ee

We use the Fourier decomposition of the fields:
\be
\tilde{A}_{k\ell}^{\mu}=\frac{1}{\sqrt{N_uN_v}}\sum_{p,q}A_{pq}^{\mu}
\exp\left[2\pi i\left(\frac{pk}{N_v}+\frac{q\ell}{N_u}\right)\right]\, ,
\ee
\bea
\tilde{\chi}_{k\ell}&=&\frac{1}{\sqrt{N_uN_v}}\sum_{p,q}\chi_{pq}
\exp\left[2\pi i\left(\frac{pk}{N_v}+\frac{q\ell}{N_u}\right)\right]\, ,\\
\tilde{\sigma}_{k\ell}&=&\frac{1}{\sqrt{N_uN_v}}\sum_{p,q}\sigma_{pq}
\exp\left[2\pi i\left(\frac{pk}{N_v}+\frac{q\ell}{N_u}\right)\right]\, .
\eea
Substituting these component fields into the lagrangian (\ref{lag}), 
we find that the photon indicated by
$A_{pq}^{\mu}$ acquire the mass:
\be
M_{pq}^2=4\left[f_u^2\sin^2\left(\frac{\pi q}{N_u}\right)
+f_v^2\sin^2\left(\frac{\pi p}{N_v}+\frac{\pi qh}{N_u}\right)\right]\, ,
\ee
while only one pseudo Nambu-Goldstone boson (PNGB), which is the combination of
$\chi_{00}$ and $\sigma_{00}$ survive.

For small $p/N_v$ and $q/N_u$, this mass spectrum becomes a continuum
KK spectrum given by
\be
M_{pq}^2\approx \left[f_u^2\left(\frac{2\pi q}{N_u}\right)^2
+f_v^2\left(\frac{2\pi p}{N_v}+\frac{2\pi qh}{N_u}\right)^2\right]\, .
\ee
This spectrum precisely corresponds to the equation (7) of Dienes'
paper\cite{Dienes}:
\be
M_{pq}^2\approx \frac{4\pi^2}{{\cal V}}\frac{1}{\tau_2}
\left[\left(p-q\tau_1\right)^2+q^2\tau_2^2 \right]\, ,
\ee
where
\be
{\cal V}=\frac{N_uN_v}{f_uf_v}\, ,\quad
\tau_2=\frac{f_uN_v}{f_vN_u}\, ,\quad
\tau_1=-\frac{N_v}{N_u}h\, .
\label{torus}
\ee

\begin{figure}[htb]
\centering
\mbox{\epsfbox{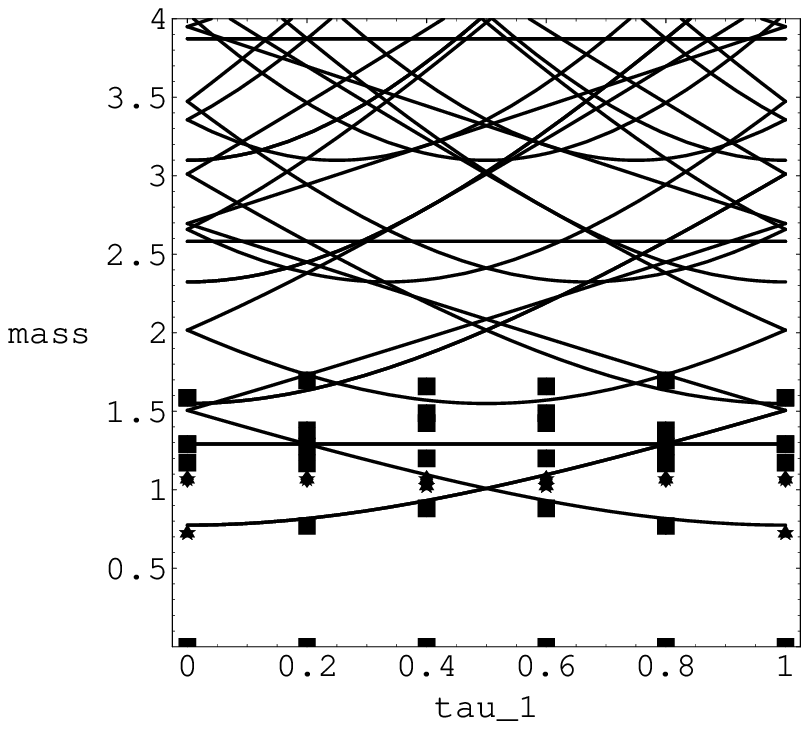}}\\
\mbox{(a)}\\
\mbox{\epsfbox{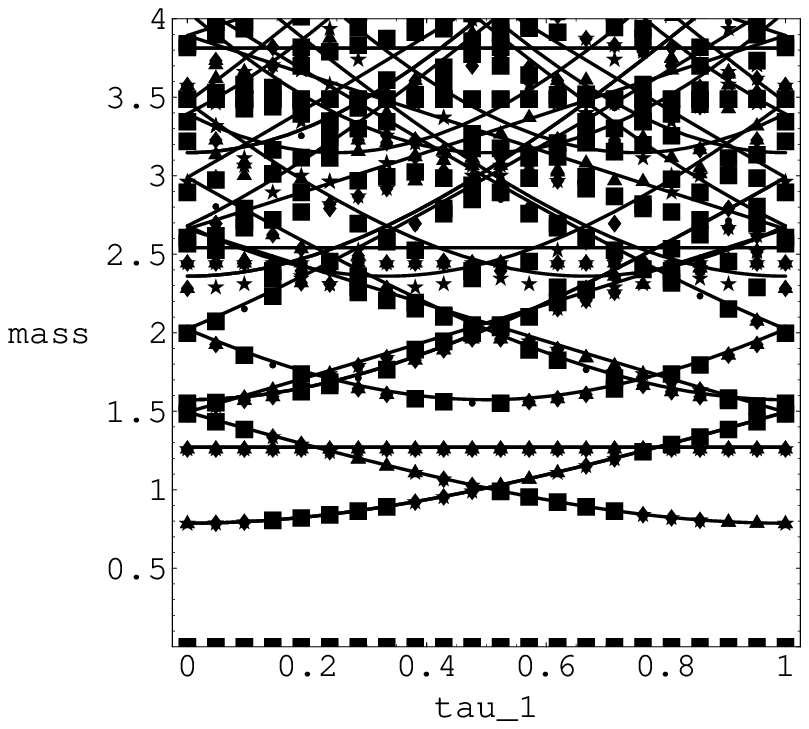}}\\
\mbox{(b)}\\
\bigskip
\caption{%
The mass spectra are plotted against the fractional part of $\tau_1$ (see text).
$(a)$ for $N_u=5$ and $N_v=3$, $(b)$ for $N_u=21$ and $N_v=13$.
Not all mass levels are exhibited in this case.}
\label{fig2}
\end{figure}

How much difference is caused when $N_u$ and $N_v$ are finite?
For example, the spectra for $(N_u,N_v)=(5,3)$ and $(21,13)$ are shown in
FIG.~\ref{fig2}. Here the horizontal axis indicates 
the fractional part of $\tau_1$.%
\footnote{For the symmetry of the figure, the same spectrum is displayed at
both $\tau_1=0$ and $\tau_1=1$.}
The vertical axis indicates $M_{pq}\sqrt{{\cal V}}/(2\pi)$.
The parameters are chosen by Eq.~(\ref{torus}) and we take $f_u=f_v$ for
simplicity.
The curves indicate the mass spectra in the large $N_u, N_v$ limit, or in the
corresponding  continuum case.

One can see that the lowest (other than zero) mass level is almost
on the curve of the continuum theory.
For $N_u=21$ and $N_v=13$, the higher levels also become close spectra 
to that in the continuum limit.
Thus it is difficult to find the discreteness of the extra space
only by the low mass spectrum, even if the torus is shifted or not.

\section{The effective potential}
\label{sec:3}

We consider the one-loop effect of charged scalar fields and assume
that $N_u\ge 3$ and $N_v\ge 3$.
The lagrangian for a complex scalar fields in our theory space
is written by
\bea
{\cal L}_{\phi}&=&
\sum_{k=1}^{N_v}\sum_{\ell=1}^{N_u}
\left[-(D^{\mu}\tilde{\phi}_{k\ell})^{\dagger}
D_{\mu}\tilde{\phi}_{k\ell}\right]\nn
&+&f_u\sum_{k=1}^{N_v}\sum_{\ell=1}^{N_u}
\left(\sqrt{2}\tilde{\phi}_{k\ell}^*U_{k\ell}\tilde{\phi}_{k,\ell+1}+
\sqrt{2}\tilde{\phi}_{k\ell}^*U_{k,\ell-1}^*\tilde{\phi}_{k,\ell-1}-
2f_u\tilde{\phi}_{k\ell}^*\tilde{\phi}_{k\ell}\right) \nn
&+&f_v\sum_{k=1}^{N_v}\sum_{\ell=1}^{N_u}
\left(\sqrt{2}\tilde{\phi}_{k\ell}^*V_{k\ell}\tilde{\phi}_{k+1,\ell+h}+
\sqrt{2}\tilde{\phi}_{k\ell}^*V_{k-1,\ell-h}^*\tilde{\phi}_{k-1,\ell-h}-
2f_v\tilde{\phi}_{k\ell}^*\tilde{\phi}_{k\ell}\right)\, ,
\eea
where
\be
D^{\mu}\tilde{\phi}_{k\ell}\equiv\partial^{\mu}\tilde{\phi}_{k\ell}-
i\tilde{A}_{k\ell}^{\mu}\tilde{\phi}_{k\ell}\, .
\ee

The scalar field can be expanded in the same manner as the gauge field:
\be
\tilde{\phi}_{k\ell}=\frac{1}{\sqrt{N_uN_v}}\sum_{p,q}\phi_{pq}
\exp\left[2\pi i\left(\frac{pk}{N_v}+\frac{q\ell}{N_u}\right)\right]\, .
\ee
Only the zero-mode fields $\chi_{00}$ and $\sigma_{00}$ affect on
the mass spectrum induced from the scalar field.
The masses are:
\be
M_{pq}^2=4\left[f_u^2\sin^2\left(\frac{\pi q}{N_u}+
\frac{\bar{\chi}}{2f_u}\right)
+f_v^2\sin^2\left(\frac{\pi p}{N_v}+\frac{\pi qh}{N_u}+
\frac{\bar{\sigma}}{2f_v}\right)\right]\, ,
\ee
where $\bar{\chi}\equiv\chi_{00}/\sqrt{N_uN_v}$ and
$\bar{\sigma}\equiv\sigma_{00}/\sqrt{N_uN_v}$.

The one-loop effective potential for $\bar{\chi}$ and
$\bar{\sigma}$ is obtained by
\bea
& &\ln\det[-\nabla^2+M^2_{pq}]\nn
&=&\lim_{\epsilon\rightarrow 0}~-\frac{1}{(2\pi)^{4-2\epsilon}}\sum_{p,q}
\int_0^{\infty}\frac{dt}{t}~
\int d^{4-2\epsilon}\Bk~\exp\left[-(\Bk^2+M_{pq}^2)t\right]\nn
&=&\lim_{\epsilon\rightarrow 0}~-\frac{1}{(4\pi)^{2-\epsilon}}\int_0^{\infty}
\frac{dt}{t}t^{-2+\epsilon}~
\sum_{p,q}\exp\left[-M_{pq}^2t\right]\, ,
\eea
after an appropriate regularization.
Using the formula\cite{GR}
\bea
\exp\left[-4f^2\sin^2(\theta/2)t\right]&=&e^{-2f^2t}\sum_{n=-\infty}^{\infty}
\cos n\theta I_n(2f^2t)\nn
&=&e^{-2f^2t}\sum_{n=-\infty}^{\infty}
e^{in\theta} I_n(2f^2t)\, ,
\eea
where $I_{\nu}(x)$ is the modified Bessel function,
we can rewrite the effective potential as
\be
V_{eff}(\bar{\chi},\bar{\sigma})=-\frac{1}{(4\pi)^{2}}\sum_{p,q}\left(\sum_{n_u,n_v=-\infty}^{\infty}\right)'
e^{in_u\theta_u+in_v\theta_v}I(n_u,n_v)\, ,
\ee
where
\be
\theta_u\equiv\frac{2\pi q}{N_u}+\frac{\bar{\chi}}{f_u}\, ,\qquad
\theta_v\equiv\frac{2\pi p}{N_v}+\frac{2\pi qh}{N_u}+
\frac{\bar{\sigma}}{f_v}
\ee
and
\be
I(n_u,n_v)=\int_0^{\infty}\frac{dt}{t^{3}}
e^{-2f_u^2t-2f_v^2t} I_{n_u}(2f_u^2t)I_{n_v}(2f_v^2t)\, ,
\label{int}
\ee
and in the summation $(\sum)'$, $n_u=n_v=0$ is omitted since
this term makes no dependence on $\bar{\chi}$ or $\bar{\sigma}$.

Carrying out the summation over $p$ and $q$,
we find that every term satisfying $n_u\frac{q}{N_u}+n_v\left(\frac{p}{N_v}
+\frac{qh}{N_u}\right)\in {\bf Z}$ is left.
Therefore we find
\bea
V_{eff}(\bar{\chi},\bar{\sigma})&=&
-\frac{N_uN_v}{(4\pi)^{2}}\left(\sum_{p',q'=-\infty}^{\infty}\right)'
\cos\left[(q'-\frac{N_v}{N_u}h~p')\frac{N_u \bar{\chi}}{f_u}+
p'\frac{N_v\bar{\sigma}}{f_v}\right]\nn
& &~~\times ~I(q'N_u-p'N_vh\,,p'N_v)\, .
\eea
Its global minima are located at
\be
\frac{N_u}{f_u}\bar{\chi}=2\pi q\, , \quad
\frac{N_v}{f_v}\bar{\sigma}=2\pi \left(p+\frac{N_v}{N_u}hq\right)\, \quad
(p, q~{\rm are~integers})\, .
\ee
These minima are of course gauge equivalent, because simplest nontrivial
 Wilson loop elements  
\be
\exp\left[i\frac{N_u\bar{\chi}}{f_u}\right]\, ,\quad
\exp\left[iN_v\left(\frac{\bar{\sigma}}{f_v}-
h\frac{\bar{\chi}}{f_u}\right)\right]\, ,
\ee
become identity at the minima of the effective potential. 

An exact expression for $I(n_u,n_v)$ is complicated as\cite{GR}
\bea
& &I(n_u,n_v)\nn
&=&
\frac{16f_v^4}{\sqrt{\pi}}\left(\frac{f_u^2}{4f_v^2}\right)^{|n_u|}
\frac{\Gamma(5/2-|n_u|)\Gamma(|n_u|+|n_v|-2)}%
{\Gamma(|n_v|-|n_u|+3)\Gamma(|n_u|+1)}\nn
& &\times {}_3F_2\left(|n_u|-|n_v|-2,|n_u|+|n_v|-2,|n_u|+1/2;
|n_u|-3/2,2|n_u|+1;-\frac{f_u^2}{f_v^2}\right)\nn &+&
\frac{32}{\pi}\frac{f_u^5}{f_v}
\frac{\Gamma(|n_u|-5/2)}{\Gamma(|n_u|-1/2)}~
{}_3F_2\left(\frac{1}{2}-|n_v|,\frac{1}{2}+|n_v|,3;
|n_u|-\frac{1}{2},-|n_u|-\frac{1}{2};-\frac{f_u^2}{f_v^2}\right)\, .
\eea
For large $n_u$ and $n_v$, the integration in (\ref{int}) can be approximated by
using the following evaluation:
\bea
I_{\nu}(z)&\sim& \frac{e^z}{\sqrt{2\pi z}}\sum_{n=0}^{\infty}
\frac{(-1)^n
\Gamma(\nu+n+\frac{1}{2})}{n!\Gamma(\nu-n+\frac{1}{2})(2z)^n}\sim
\frac{e^z}{\sqrt{2\pi
z}}\sum_{n=0}^{\infty}
\frac{(-1)^n \nu^{2n}}{n!(2z)^n}\nn
&\sim&\frac{e^z}{\sqrt{2\pi z}}\exp(-\frac{\nu^2}{2z})\qquad(z\gg 1)\, .
\eea
Then the integration results in a rather simple form for large $n_u$ and $n_v$:
\be
I(n_u,n_v)=\frac{32}{\pi}\frac{1}{f_uf_v}
\frac{1}{(n_u^2/f_u^2+n_v^2/f_v^2)^3}\, .
\ee
By using this with the notation (\ref{torus}), we find that the effective potential
in the large $N_u$ and $N_v$ limit,
\be
V_{eff}(\bar{\chi},\bar{\sigma})=
-\frac{2}{\pi^{3}}\frac{1}{{\cal V}^2}\left(\sum_{p',q'=-\infty}^{\infty}\right)'
\frac{\cos\left[(q'+p'\tau_1)\frac{N_u\bar{\chi}}{f_u}+
p'\frac{N_v\bar{\sigma}}{f_v}\right]}%
{[(q'+p'\tau_1)^2/\tau_2+(p')^2\tau_2]^3}\, ,
\ee
is that in the continuum $U(1)$ gauge theory with compactification
on the shifted two-torus.

When the fractional part of $\tau_1$ does not vanish,
the mass matrix of PNGBs has off-diagonal parts for any $N$.

\section{conclusion and discussion}
\label{sec:4}

In conclusion, we can build a theory space description of the shifted-torus
compactification.
The correspondence in the limit of large number of sites is found
when the volume of extra space held fixed.

Although the resulting mass spectra contain a finite number of
massive states in our models, the lowest non-zero mass is very close to that of
the corresponding continuum models.
Therefore it is difficult to distinguish the discrete 
and continuum theories experimentally through the direct production of
the lowest KK state.
If we can observe the virtual exchange contribution of
the KK states to cross sections, we can study the mass spectrum
experimentally. For this purpose, we must compute amplitudes
along with more realistic theories. This subject is left for
future works.

We have also discussed the effective potential for PNGBs in our shifted-lattice
model. We will perform more quantitative analysis of the effective potential for
simple models of deconstruction. Moreover, 
the possibility of finite temperature effect on the effective potential
will be discussed elsewhere\cite{fw}.

\begin{acknowledgments}
We would like to thank Y. Cho for his valuable comments
and for the reading the manuscript.
\end{acknowledgments}



\end{document}